# On Modelling The Immune System as a Complex system

E. Ahmed* and A.H. Hashish**

*Department of Mathematics, Faculty of Science, Mansoura Univ. Mansoura 35516, EGYPT.
** Department of Physics**, Faculty of science, UAE University, Al-Ain, P.O.Box 17551, United Arab Emirates.

{Abstract:}
  We argue that immune system is an adaptive complex system. It is shown that it has emergent properties. Its network structure is of the small world network type. The network is of the threshold type, which helps in avoiding autoimmunity. It has the property that every antigen (e.g.virus or bacteria) is typically attacked by more than one effector. This stabilizes the equilibrium state. Modelling complex systems is discussed. Cellular automata (CA)\ type models are successful but there are much less analytic results about CA than about other less successful models e.g. partial differential equations (PDE). A compromise is proposed

{1. Basic definitions:}
  The immune system (IS) is one of the most advanced and complex biological systems. It is a complex network of cells and chemicals. Its mission is to protect us against foreign organisms and substances. The cells in the IS have the ability to recognize objects as either damaging or nondamaging. Many different kinds of cells, and hundreds of different chemicals, must be coordinated for the IS to function successfully.
  IS consists of master glands, principally the thymus and the bone marrow; various sites that temporarily harbor immune cells; and different classes of "soldier" cells, which carry out specialized functions, including cells that alert, cells and molecules that facilitate, cells that activate, cells that engulf, cells that kill, even cells that clean up. Many immune cells also synthesize and secrete special molecules that act as messengers, regulators, helpers or suppressors in the process of defending against invaders.
  Like the nervous system, the IS performs pattern recognition tasks, learns and retains a memory of the antigens that it has fought. The IS contains more than 100 000 000 000 different clones of cells that communicate via cell-cell contact and the secretion of molecules. Performing complex tasks such as learning and memory involves cooperatively among large numbers of components of the IS and hence there is interest in using the methods and the concepts of statistical mechanics to understand such a complex system.
  We start by presenting a basic definition of a complex system:
Definition (1): "A complex system, [Smith, 2003], is a one with multiple interacting elements whose collective behavior cannot be simply inferred from the behavior of its elements".

This totalistic approach is against the standard reductionist one, which tries to decompose any system to its constituents and hopes that by understanding the elements one can understand the whole system.

Complex systems are abundant in nature e.g. the brain, insects' swarms, etc. Here we argue that IS, [Roitt and Delves, 2000] ,falls into this category.

The main components of IS are the innate and the adaptive ones. The innate component contains (among others) the complement system, macrophage, dendrite cells, and natural killers. The adaptive component contains the B-cells, Th-cells and Tc-cells. Both components are interconnected by a large number of cytokines that cause activation and/or inhibition. This forms the immune network.

Definition (2): "An emergent property of a complex system is a property of the complex system as a whole but it does not exist in each individual element".

The emergent properties of IS included:
1) The ability to distinguish any substance (typically called antigen Ag) and determine whether it is self or nonself.
2) If Ag is nonself then IS determines whether to tolerate it or to respond to it.
3) If it decides to respond to it then IS determines whether to eradicate or to contain it.
4) The ability to memorize most previously encountered Ag, which enables it to mount a more effective reaction in any future encounters. This is the basis of vaccination processes.

{2. The immune network:}

The immune network interactions are very complicated and far from being well understood but a naive description is as follows:

When a foreign humoral Ag (e.g. bacteria) invades the body dendrite cells (and other Ag presenting cells (APC)) identify it, process it and present it to the Th-cells together with a second signal to stimulate the Th-cells.

Without this second signal the Th-cells go into a state of anergy (become paralyzed). Once activated the Th-cells activates (through cytokines) B-cells and macrophages.

B-cells secrete antibodies (Ab), which together with the complement system and macrophage eliminate the Ag. While this process is going on, some B-cells become memory B-cells. Such cells are long lived that can mount a faster and stronger effect against that Ag in future encounters.

A similar scenario occurs for cellular Ag (e.g. viruses) but in this case Th activates Tc, natural killers and macrophage to attack infected cells and eradicate the virus. While this process is going on, some T-cells become memory T-cells

This shows that the immune network is neither homogeneous, nor all to all, nor random. Rather there are elements, e.g. Th, which are better connected than others. Therefore immune network is topologically a small world network [Watts and Strogatz, 1998]. This kind of networks is expected to be abundant in nature. They are strongly resistant to random errors but weak against errors, which affect highly connected elements. This is a reason why AIDS is quite dangerous since it attacks the highly connected Th cells. Thus it disrupts the immune network making the individual vulnerable to opportunistic Ag.

The Ag, which enters our bodies, has extremely wide diversity. Thus mechanisms have to exist to produce immune effectors with constantly changing random specificity to be able to recognize these Ag. Consequently IS is an adaptive complex system. Having said that, one should notice that wide diversity of IS contains the danger of autoimmunity (attacking self). Thus mechanisms that limit autoimmunity should exist. In addition to the primary clonally deletion mechanism, two further brilliant mechanisms exist: The first is that the IS network is a threshold one i.e. no activation exists if the Ag quantity is too low or too high (This is called low and high zone tolerance). Thus a self-reactive immune effector (i.e. an immune effector that attacks the body to which it belongs) will face so many self-antigens that it has to be suppressed due to the high zone tolerance mechanism. Another mechanism is the second signal given by APC. If the immune effector is self then, in most cases, it does not receive the second signal thus it becomes anergic.

Finally to get a better understanding of IS the following points may be relevant:

a) IS functions in the presence of noise, hence exact identification is not possible. Some fuzziness seems inevitable.

b) IS effectors are multi-function, i.e. most of them have more than one function to perform (e.g. macrophage both engulf Ag and presents Ag to activate mature Th-cells). Also almost every function of IS is done by more than one effector (IS is multipathways). Obviously this ensures better success in Ag identification and elimination. It also supports the idea that IS is a complex system. This multipathway property stabilizes the equilibrium state [Nowak and May 2001].

c) This multifunctionality may differ from our concepts of optimality, however, we should remember that in our problems we typically have a good idea about the problem we are facing. This is not the case for IS which faces almost arbitrary Ag.

d) There are similarities between IS and social insects colonies, e.g. ants and bees [Segel and Cohen, 2001]. This may be helpful in getting better understanding of IS network.

e) IS is a decentralized system.

f) IS has multi-objectives since it has to maximize harmful Ag elimination and at the same time minimize harm to self (autoimmunity). All this should be done with limited resources.

{3.Modelling complex systems}

Having shown that IS is a complex system a question now arises as to how to model a complex system? It has been argued [Louzon et al 2003] that neither ODE nor partial differential equations PDE are quite suitable to model complex biological systems for the following reasons:

1) Both ODE and PDE assumes that local fluctuations have been smoothed out.
2) Typically they neglect correlations between movements of different species.
3) They assume instantaneous results of interactions. Most biological systems (including IS) shows delay and do not satisfy the above assumptions. They concluded that a cellular automata (CA) [Ilachinski 2001] type system called microscopic simulation is more suitable to model complex biological systems. We agree that CA type systems are more suitable to model complex biological systems but such systems suffer from a main drawback namely the difficulty of obtaining analytical results. The known analytical results about CA type systems is very few compared to the known results about ODE and PDE.

Therefore we present a compromise i.e. a PDE which avoids the delay and the correlations drawbacks. It is called telegraph reaction diffusion equations [Ahmed and Hassan 2000]. To overcome the non-delay weakness in Fick's law it is replaced by

$$J(x, t) + \tau \partial J/\partial t = -D \partial c/\partial x \qquad (1)$$

where now the flux relaxes, with some given characteristic time constant $\tau$.
Combining (1) with the equation of continuity one obtains the modified diffusion equation or (Telegraph equation)

$$\partial c/\partial t + \tau \partial^2 c/\partial t^2 = D\partial^2 c/\partial x^2 \quad (2)$$

The corresponding Telegraph reaction diffusion (TRD) is given by

$$(1 - \tau df/dc)\partial c/\partial t + \tau \partial^2 c/\partial t^2 = D\partial^2 c/\partial x^2 + f(c) \quad (3)$$

the time constant $\tau$ can be related to the memory effect of the flux J as a function of the distribution c as shown in what follows.

Another motive for TRD comes from media with memory where the flux J is related to the density c through a relaxation function K(t) as follows:

$$J(x,t) = -\int_0^t K(t-s)\partial c(x,s)/\partial x \, ds \quad (4)$$

we will see that, with a suitable choice for K(t), the standard Telegraph equation is obtained, Indeed, let us compute the left-hand side of the equation (4). For our generalization we get:

$$J + \tau \partial J/\partial t = -(\tau \partial/\partial t + 1)(\int_0^t K(t-s)\partial c(x,s)/\partial x \, ds) \quad (5)$$

Choosing $K(t) = (D/\tau)\exp(-t/\tau)$ we get (2).

Moreover it is known that TRD results from correlated random walk [Diekmann et al, 2000]. This supports that Telegraph diffusion equation is more suitable for economic and biological systems than the usual one since, e.g., it is known that we take our decisions according to our previous experiences so memory effects are quite relevant.

{Conclusions:}

It is argued that immune system is a complex adaptive system. Its emergent properties are mentioned. To model a complex system it has been argued that cellular automata type approach is quite suitable. The problem is that there are much less analytic results about CA than differential equations. Here a compromise is proposed namely telegraph reaction diffusion equation.

{Acknowledgment}

We thank the referee for helpful comments.

{References:}